\documentclass[aps,prl,reprint,groupedaddress,showkeys]{revtex4-1}

\usepackage{amsmath}
\usepackage{amssymb}
\usepackage[pdftex]{hyperref}

\renewcommand{\d}{\mathrm{d}}
\newcommand{\e}{\mathrm{e}}

\newcommand{\nl}{\notag \\ &\quad\,}

\begin{document}

\title{Dynamics of warped flux compactifications with backreacting anti-branes}

\author{Daniel Junghans}
\email[]{daniel@ust.hk}

\affiliation{Center for Fundamental Physics \& Institute for Advanced Study, The Hong Kong University of Science and Technology, Hong Kong}

\date{\today}

\begin{abstract}
We revisit the effective low-energy dynamics of the volume modulus in warped flux compactifications with anti-D$3$-branes in order to analyze the prospects for meta-stable de Sitter vacua and brane inflation along the lines of KKLT/KKLMMT. At the level of the 10d supergravity solution, anti-branes in flux backgrounds with opposite charge are known to source singular terms in the energy densities of the bulk fluxes, which led to a debate on the consistency of such constructions in string theory. A straightforward yet non-trivial check of the singular solution is to verify that its dimensional reduction in the large-volume limit reproduces the 4d low-energy dynamics expected from known results where the anti-branes are treated as a probe.
Taking into account the anti-brane backreaction in the effective scalar potential, we find that both the volume scaling and the coefficient of the anti-brane uplift term are in exact agreement with the probe potential if the singular fluxes satisfy a certain near-brane boundary condition.
This condition can be tested explicitly and may thus help to decide whether flux singularities should be interpreted as pathological or benign features of flux compactifications with anti-branes. Throughout the paper, we also comment on a number of subtleties related to the proper definition of warped effective field theory with anti-branes.
\end{abstract}

\pacs{}

\maketitle

\section{I. Introduction}

Inserting anti-D$3$-branes into warped throat regions of flux compactifications was proposed as a mechanism to construct meta-stable de Sitter vacua in string theory \cite{Kachru:2003aw} (KKLT) and to realize brane inflation \cite{Dvali:1998pa} with sufficiently flat potentials to allow for sustained slow-roll \cite{Kachru:2003sx} (KKLMMT). In \cite{Kachru:2003aw}, it was argued that the low-energy dynamics of a large class of string compactifications realizing this mechanism are captured by a 4d effective scalar potential of the form
\begin{equation}
V(\rho) = V_\textrm{np}(\rho) + V_{\overline{\textrm{D}3}}(\rho), \label{potential}
\end{equation}
where $\rho$ is the volume or universal K{\"{a}}hler modulus. To set our notation, let us quickly review the two terms in this potential (for more details of the construction, we refer to \cite{Kachru:2003aw}). The first term on the right-hand side of \eqref{potential} can be written as
\begin{align}
V_\textrm{np}(\rho) &=  \frac{1}{\rho^2} \left({W^\prime (\rho)^2 \rho - 3W(\rho)W^\prime(\rho)}\right) \label{potential-np}
\end{align}
in terms of a superpotential
\begin{equation}
W(\rho) = W_0 + W_\textrm{np}(\rho), \qquad W_\textrm{np}(\rho) = A \e^{-a\rho} \label{superpotential-np}
\end{equation}
in the language of $\mathcal{N}=1$ supergravity. Here, $W_0$ is a term generated by fluxes at tree-level \cite{Gukov:1999ya}, which are chosen such that they carry D$3$-brane charge, and $W_\textrm{np}$ is due to non-perturbative instanton corrections with model-dependent coefficients $A$ and $a$ \cite{Witten:1996bn, Kachru:2003aw}.

In the probe approximation, the anti-D$3$-branes can be shown to contribute to \eqref{potential} with a term
\begin{align}
V_{\overline{\textrm{D}3}}(\rho) &= \frac{D}{\rho^\gamma}, \qquad D=2 p T_3 \e^{4A_\textrm{bg}(y_0)}, \label{potential-d3}
\end{align}
which explicitly breaks supersymmetry \cite{Kachru:2003aw}. Here, $D$ is a coefficient that depends on the tension $T_3$ and the number $p$ of the anti-branes and on the background warp factor $\e^{4A_\textrm{bg}(y_0)}$ at the anti-brane position $y_0$. The parameter $\gamma$ determines the scaling of the uplift term with respect to the volume modulus and was argued to equal $3$ for weakly warped backgrounds and $2$ in the case of strong warping \cite{Kachru:2003aw, Kachru:2003sx}. By ``probe approximation'' we mean that the backreaction of the anti-branes is considered only in an integrated sense, i.e., their charge is accounted for in the tadpole condition but the effect of the non-trivial brane profile on the internal fields is assumed to be negligible from the point of view of the 4d low-energy dynamics. This is expected to be justified in the limit of a large internal volume and a small number of anti-branes. From the 10d point of view, it can be thought of as ``smearing'' the brane profile over the compact space such that it locally cancels the extra flux needed for tadpole cancellation.
Accordingly, $\e^{4A_\textrm{bg}}$ refers to the warp factor that would be generated by a flux background consistent with the presence of $p$ smeared anti-branes.

By tuning the superpotential and the background warp factor in \eqref{potential-np} and \eqref{potential-d3}, $\rho$ can be stabilized at large volume and small, positive vacuum energy, yielding a meta-stable de Sitter vacuum \cite{Kachru:2003aw}. Furthermore, brane inflation can be realized by inserting a D$3$-brane into such a background, where the inflaton is identified with the position of the brane as it moves towards the anti-brane. The stabilization of the volume modulus in this scenario generically leads to a large contribution to the $\eta$ parameter \cite{Kachru:2003sx}, however, it was argued that this term can be cancelled by various compactification effects such that slow-roll inflation is still possible in certain regions of the landscape \cite{Berg:2004ek, Baumann:2006th, Baumann:2007np, Baumann:2007ah, Krause:2007jk, Baumann:2008kq, Baumann:2010sx}.

Due to its phenomenological appeal, the idea of placing anti-branes in flux backgrounds with opposite charge has received a lot of attention in the last years. In particular, much work has been devoted to constructing explicit flux compactifications realizing this idea.
For concreteness, we will focus on anti-D$3$-branes that are inserted into the Klebanov-Strassler (KS) background \cite{Klebanov:2000hb}, a well-understood example for a strongly warped throat that can be embedded into a compact setting \cite{Giddings:2001yu}. The backreaction of the anti-D$3$-branes on the KS background has been heavily studied in recent years \cite{DeWolfe:2008zy, McGuirk:2009xx, Bena:2009xk, Bena:2011hz, Bena:2011wh, Dymarsky:2011pm, Massai:2012jn, Bena:2012bk, Gautason:2013}, which led to the discovery of (integrable) singularities in the energy densities of the NSNS and RR $3$-form field strengths $H$ and $F_3$ at the anti-brane position. Under the simplifying assumption of partially smeared anti-branes, this was first observed in \cite{McGuirk:2009xx, Bena:2009xk} at linear order in a perturbation around the BPS background and later in \cite{Bena:2012bk} taking into account the full non-linear backreaction. Recently, it was then shown that the singularities are also generated by the backreaction of fully localized anti-branes \cite{Gautason:2013}. The interpretation of the flux singularities has been subject of an intense debate, but so far no consensus has been reached about whether or not they are acceptable in string theory (see \cite{Blaback:2012nf, Bena:2012tx, Bena:2012vz, Bena:2012ek, Vanriet:2013, Buchel:2013dla} for counter-arguments).

The aim of this note is to add a new perspective to this discussion, i.e., to examine the effects of the anti-brane backreaction from the point of view of the 4d warped effective field theory \cite{DeWolfe:2002nn, Giddings:2005ff, Frey:2006wv, Burgess:2006mn, Douglas:2007tu, Koerber:2007xk, Shiu:2008ry, Douglas:2008jx, Frey:2008xw, Marchesano:2008rg, Martucci:2009sf, Chen:2009zi, Douglas:2009zn, Underwood:2010pm, Marchesano:2010bs, Blaback:2012mu, Frey:2013bha} that governs the low-energy dynamics of the compactification. In particular, we are interested in how the anti-brane backreaction affects the uplift term \eqref{potential-d3}, a question that is largely unexplored in the literature (see \cite{Douglas:2010rt, Blaback:2010sj, Blaback:2011nz} for a general discussion of the smeared approximation).
In this note, we will take some initial steps in this direction. Since the solution for anti-branes in the KS background is only partially known, we cannot be fully explicit and necessarily have to rely on some assumptions that we will comment on later. Furthermore, we will restrict ourselves to an analysis of the leading order potential and leave the computation of sub-leading corrections for future work.

Our main observation is that, in the large-volume limit and under some additional assumptions, the probe potential \eqref{potential-d3} exactly agrees with the potential one obtains taking into account the full backreaction of the anti-branes. More precisely, both the coefficient $D$ and the scaling $\gamma$ of the uplift term match, provided a certain boundary condition at the anti-brane position is satisfied. This is quite remarkable since, in the 10d supergravity solution, the singular fluxes scale with a higher power of the inverse warp factor than regular fluxes would. Since the warp factor is related to the volume modulus in the warped effective field theory, one might have expected that the backreaction of the singular solution is too large and that only a (hypothetical) solution without flux singularities could have reproduced the correct scaling behavior of \eqref{potential-d3}. We will argue below, however, that this is not the case if above mentioned boundary condition is satisfied.

The condition may therefore be used as a consistency check of the singular supergravity solution: if it is violated in the appropriate limit, the singular solution does not reproduce the potential expected from a probe analysis, which suggests that it is unphysical. If, on the other hand, the condition is satisfied, the singular solution yields the correct low-energy dynamics, which may in turn be taken as an indication (although it is of course no proof) that it is a consistent solution in string theory. If this is true, flux singularities are resolved by some (yet unknown) stringy mechanism and should consequently not be interpreted as pathological but as a harmless feature of the supergravity solution.

To conclude this section, let us point out that the appearance of flux singularities is not specific to anti-D$3$-branes in the KS background but in fact a much more general phenomenon. Flux singularities were discovered in setups with anti-D2-branes \cite{Giecold:2011gw}, anti-M2-branes \cite{Bena:2010gs, Massai:2011vi, Giecold:2013pza, Cottrell:2013asa, Blaback:2013hqa, Bena:2014bxa} and anti-D6-branes \cite{Blaback:2011nz, Blaback:2011pn, Apruzzi:2013yva} in various flux backgrounds with opposite charge. In \cite{Gautason:2013}, it was understood from a more general point of view why these singularities arise. There, they were shown to originate from global constraints that relate the boundary conditions at the anti-brane position to the cosmological constant (or, in a non-compact setting, to the boundary conditions far away from the anti-branes, e.g., at the end of a warped throat). These constraints are due to the specific way of how branes and fluxes break the classical scale invariance of the supergravity equations (see also \cite{Aghababaie:2003ar, Burgess:2011rv} for earlier works exploiting scaling symmetries) and insensitive to local details of the particular solution. Hence, the appearance of flux singularities is a universal property of a large class of supergravity solutions. It is possible that they also arise in string compactifications with D-term uplifting, where a positive contribution to the vacuum energy is generated by turning on worldvolume fluxes on D$7$-branes \cite{Burgess:2003ic}. The latter then carry a negative D$3$-brane charge and thus act like partially smeared anti-D$3$-branes. We therefore expect that the discussion of this paper is not only useful in the context of the KKLT scenario but applicable to a much wider range of models, for anti-branes of different dimension and for other flux backgrounds.

\section{II. Effective scalar potential}

Let us now consider a type IIB compactification for which the complex structure moduli and the dilaton are stabilized by fluxes at tree-level as in \cite{Giddings:2001yu} (see also \cite{Becker:1996gj,Dasgupta:1999ss,Gukov:1999ya,Greene:2000gh} for earlier work). At low energies, we can then integrate out these degrees of freedom and consider an effective action for the K{\"{a}}hler moduli as usual.
To avoid an unnecessary complication of our discussion, we will assume a model with only one K{\"{a}}hler modulus, i.e., the volume (see \cite{Shiu:2008ry, Frey:2013bha} for analyses of non-universal K{\"{a}}hler moduli in warped effective field theory). It was argued that achieving moduli stabilization by non-perturbative effects can be difficult in such models \cite{Denef:2004dm, Robbins:2004hx} (see, however, \cite{Bobkov:2010rf, Bianchi:2011qh}). While this issue must be addressed in a fully explicit model, our main interest in this paper is in how the anti-brane backreaction affects the uplift term. We will therefore assume a non-perturbative superpotential as in \eqref{superpotential-np} without discussing its explicit realization. Furthermore, we will assume that, as argued originally in \cite{Kachru:2003aw}, the non-perturbative effects do not significantly backreact on the classical supergravity background such that their presence is fully captured by adding the term \eqref{potential-np} to the scalar potential (see also \cite{Gautason:2013} for a more detailed discussion). For analyses of the backreaction of non-perturbative effects, we refer to \cite{Koerber:2007xk, Baumann:2010sx, Heidenreich:2010ad, Dymarsky:2010mf}.
For the sake of explicitness, let us finally consider a model where D$3$ tadpole cancellation is ensured via the presence of O$3$-planes in the bulk \cite{Giddings:2001yu}. Our conclusions should be unaffected by this choice, and we expect that analogous results hold if the KS solution is embedded into more general compactifications such as the F-theory models discussed in \cite{Giddings:2001yu}.

The metric can be written in the form
\begin{equation}
\d s_{10}^2 = \e^{2A(y)} \tilde g_{\mu\nu} \d x^\mu \d x^\nu + \e^{-2A(y)} \tilde g_{mn} \d y^m \d y^n, \label{metric}
\end{equation}
where we have split off a warp factor $\e^{2A}$ from the external part of the metric and a conformal factor $\e^{-2A}$ from its internal part. Before the insertion of the anti-D$3$-branes, $\tilde g_{mn}$ is known to be Ricci-flat \cite{Giddings:2001yu}, and, close to the O$3$-planes and the D$3$-branes (if present), the internal metric diverges like $\e^{-2A}$. Furthermore, we expect from the well-known brane solutions in flat space \cite{Ortin:2004ms} that, close to an anti-D$3$-brane, the internal metric also diverges like $\e^{-2A}$ (i.e., $\tilde g_{mn}$ does not have any curvature singularities at the anti-brane position). That this is indeed true in the KS background was explicitly confirmed in \cite{Bena:2012vz} in a simplified setup with partially smeared anti-branes, and we will assume in the following that it remains true for fully localized anti-branes. The conformal factor in \eqref{metric} then captures the expected divergence behavior close to all localized sources even after the insertion of the anti-branes.
Note, however, that $\tilde g_{mn}$ need then not be Ricci-flat anymore but will in general be deformed away from a Calabi-Yau.

{It was recently argued in \cite{Bena:2014bxa} that, on top of the standard singularities familiar from brane solutions in flat space, the warp factor and/or the conformal factor in \eqref{metric} might develop even stronger singularities at the anti-brane position due to the backreaction of the bulk fluxes. To keep the discussion simple, we will not consider this possibility in detail, but let us note here that our arguments still go through under these more general conditions as long as the metric remains of the form \eqref{metric} everywhere except very close to the anti-branes. In particular, one can convince oneself that the main conclusion of this paper, i.e., the boundary condition proposed further below, is not affected by how strongly the warp factor diverges at the anti-brane position.}

Our goal is now to derive the 4d effective action of the volume modulus in the large-volume limit of the above described class of flux compactifications. As stated before, we will, at least at leading order, take into account the anti-brane backreaction exactly but, as in \cite{Kachru:2003aw}, treat the non-perturbative effects as an additive term in the scalar potential without attempting to derive its 10d origin. The logic of our derivation is then as follows. We start with the premise that there is a mechanism in string theory that resolves the flux singularities at short distances such that inserting anti-branes into the KS background yields a consistent compactification. Under this assumption, we will derive a boundary condition that must be satisfied at the anti-brane position in order that the low-energy dynamics of the volume modulus can be matched with what one would expect from a probe analysis.

To be more precise, let us assume that the string coupling $g_s$ can be made small and that the singular supergravity solution correctly describes the field behavior sufficiently far away from the anti-branes up to a distance $\alpha^\prime/\ell_\textrm{c}$ below which string corrections become large and resolve the singularities. In the large-volume limit, the string length $\sqrt{\alpha^\prime}$ is small compared to the length scale $\ell_\textrm{c}$ set by the compactification such that the supergravity approximation is valid almost everywhere on the compact space. The 4d effective field theory governing the low-energy dynamics of the system is therefore well-approximated by a dimensional reduction of the type IIB tree-level supergravity action, and string corrections, while important close to the anti-branes and other localized sources, are negligible in the 4d effective potential. This allows us to study the low-energy dynamics of the compactification without having to explicitly know the mechanism that resolves the flux singularities.

It was shown in \cite{Kachru:2002gs} that probe anti-D$3$-branes in the KS background can polarize into an NS$5$-brane that wraps a trivial $2$-cycle inside of the KS throat. If this effect (or, more generally, a polarization into a $(p,q)$ $5$-brane as in \cite{Polchinski:2000uf}) also occurs for backreacting anti-branes, the singular supergravity solution is valid only at distances larger than the polarization radius $\ell_\textrm{pol}$, while the delocalization of the anti-brane charge at short distances may lead to a resolution of the flux singularities (see, however, \cite{Bena:2012tx, Bena:2012vz}). Above discussion then applies accordingly if the polarization radius is much smaller than the compactification scale, $\ell_\textrm{pol}/\ell_\textrm{c} \ll 1$.

Substituting the metric ansatz \eqref{metric} into the type IIB supergravity action, performing a dimensional reduction and switching to 4d Einstein frame, we find the 4d effective action
\begin{equation}
S \supset \int \d^4 x \sqrt{-\tilde g_4}\, \left( \tilde R_4 - V \right), \label{effaction}
\end{equation}
where we have set $2\kappa_{4}^2=1$ and omitted to write down the kinetic terms for the moduli. The general form of the effective scalar potential is
\begin{align}
V &= \int \d^{6}y \sqrt{\tilde g_{6}}\, \frac{1}{\mathcal{V}_\textrm{w}^{2}} \bigg[ - \tilde R_6 + 8 (\tilde{\partial A})^2  + \frac{1}{2} (\tilde{\partial \phi})^2 \nl + \frac{1}{2} \e^{4A-\phi} |\tilde H|^2 + \frac{1}{2} \e^{4A+\phi}|\tilde F_3|^2 + \frac{1}{2}\e^{8A} |\tilde F^\textrm{int}_5|^2 \nl + \Big(p+N_{\textrm{D}3}-\frac{1}{4}N_{\textrm{O}3}\Big)\, T_3 \e^{4A} \tilde \delta^{(6)} \bigg] + V_\textrm{np}, \label{effaction2}
\end{align}
which should be read as implicitly depending on the volume modulus. Here, $\mathcal{V}_\textrm{w}=\int \d^6 y  \sqrt{\tilde g_6}\, \e^{-4A}$ is the warped volume of the internal space, and $\tilde R_4$ and $\tilde R_6$ are the Ricci scalars constructed from $\tilde g_{\mu\nu}$ and $\tilde g_{mn}$, respectively. Furthermore, tildes on top of energy densities indicate contractions with unwarped metrics. The first term in the last line is due to the DBI action of the localized sources, where $T_3$ is the tension of one D$3$-brane (which equals $-4$ times the tension of one O$3$-plane) and $p$, $N_{\textrm{D}3}$ and $N_{\textrm{O}3}$ are the numbers of the (anti-)D$3$-branes and O$3$-planes. $\tilde \delta^{(6)}$ is a formal sum of delta distributions with support at the positions of the localized sources.

Before we proceed, let us comment on two subtleties related to the definition of the scalar potential in warped effective field theory. One well-known problem is due to the self-dual $F_5$ field strength in type IIB string theory. Dimensionally reducing the 10d supergravity action and then imposing the self-duality relation in the 4d effective action leads to the wrong scalar potential \cite{DeWolfe:2002nn}.
An ad hoc recipe to obtain the correct potential is to first discard the external part of $F_5$ and double its internal part $F^\textrm{int}_5$ in the type IIB action and then perform the dimensional reduction \cite{DeWolfe:2002nn}. That this is indeed the correct procedure can be verified by computing the external part of the 10d (trace-unreversed) Einstein equations and then reading off the scalar potential using the relation $\tilde R_4 = 2 V$ (see \cite{Giddings:2005ff} for a detailed discussion).

Another issue, which was pointed out in \cite{Douglas:2009zn}, is that localized objects in warped compactifications generically source a conformal factor $\e^{2B}$ in the internal metric $g_{mn} = \e^{2B} \tilde g_{mn}$, which varies over the internal space. One can show that this induces curvature terms $\sim\!(\partial B)^2$ in the scalar potential that come with a negative sign, which has the effect that the potential for $B$ is not bounded from below and thus unstable against rapid fluctuations. The metric \eqref{metric} avoids this problem since here the conformal factor is not an independent degree of freedom but tied to the warp factor via $B=-A$. The combined potential for $A$ and $B$ is then bounded from below, as can be seen from \eqref{effaction2}. It is tempting to speculate that all consistent warped compactifications must have this property of a ``melting'' of the conformal factor with other degrees of freedom to avoid instabilities and other problems (see also \cite{Underwood:2010pm} for a related phenomenon involving the dilaton).

\section{III. Dynamics of the volume modulus}

Our next goal is to make explicit the dependence of the scalar potential \eqref{effaction2} on the volume modulus. We will first review the BPS case without any anti-D$3$-branes where the tree-level potential is flat due to a no-scale structure of the 10d supergravity solution \cite{Giddings:2001yu}. We then discuss what happens when we insert D$3$-branes and anti-D$3$-branes into such a background.
While D$3$-branes are mutually BPS with the flux background and do therefore not break the no-scale structure, inserting anti-D$3$-branes yields a positive contribution to the scalar potential that can be matched with the uplift term \eqref{potential-d3} in the appropriate limit.

\subsection{a. BPS flux compactifications}

Before inserting anti-D$3$-branes, the compactification considered in this paper is described by the BPS flux background of \cite{Giddings:2001yu} at tree-level.
The supergravity solution then satisfies the 10d equations of motion with
\begin{gather}
H = \e^{\phi-4A} \star_6 \alpha F_3, \quad F_5 = - (1+\star_{10}) \e^{-4A} \star_6 \d \alpha, \notag \\
\tilde R_6=0, \quad \phi=\textrm{const.}, \quad \alpha = \e^{4A}. \label{gkp}
\end{gather}
One can show that the solution saturates a BPS bound, which ensures its stability. Furthermore, it is of no-scale type, i.e., the volume modulus is not stabilized at tree-level and only receives a potential from the non-perturbative term $V_\textrm{np}$ (and from other string/quantum corrections, which are, however, assumed to be sub-leading in the KKLT scenario).
The tree-level equations of motion must therefore be invariant under variations of the volume modulus. Unlike in the unwarped case, this is not true for a simple rescaling of the internal metric, as this would not leave the $F_5$ Bianchi identity
\begin{equation}
\tilde \nabla^m \e^{-8A} \partial_m \alpha = \alpha\e^{-4A+\phi} |\tilde F_3|^2 - \frac{1}{4} N_{\textrm{O}3} T_3 \tilde \delta^{(6)} \label{bianchi}
\end{equation}
and the external part of the 10d Einstein equation
\begin{equation}
\tilde \nabla^m \e^{-8A} \partial_m \e^{4A} = \frac{1}{2}\left(1+\alpha^2\e^{-8A}\right) \e^\phi |\tilde F_3|^2 - \frac{1}{4} N_{\textrm{O}3} T_3 \tilde \delta^{(6)} \label{einstein}
\end{equation}
invariant \cite{Giddings:2005ff}. Both equations are, however, invariant under constant shifts $\e^{-4A(y)} \to \e^{-4A(y)} + \rho$, $\alpha(y)^{-1} \to \alpha(y)^{-1} + \rho$ (in fact, this is true for all equations of motion). The easiest way to see this is to use the BPS property $\alpha=\e^{4A}$ in \eqref{bianchi} and \eqref{einstein}. The equations then simplify to
\begin{equation}
- \tilde \nabla^2 \e^{-4A} = \e^\phi|\tilde F_3|^2 - \frac{1}{4} N_{\textrm{O}3} T_3 \tilde \delta^{(6)}, \label{bianchi3}
\end{equation}
which is manifestly invariant under shifts of the warp factor.

Hence, it is natural to identify $\rho$ with the volume modulus \cite{Giddings:2005ff}. For future convenience, let us treat $\e^{4A}$ and $\alpha$ as two separate functions as in \eqref{bianchi} and \eqref{einstein}. We can then define the fluctuating fields as functions of $\rho$,
\begin{align}
\e^{-4A(\rho(x),y)} &= \e^{-4A_0(y)} + \rho(x), \label{warped-modulus2} \\
\alpha^{-1}(\rho(x),y) &= \alpha_0^{-1}(y) + \rho(x), \label{warped-modulus3}
\end{align}
where $\e^{4A_0}$ and $\alpha_0$ denote the solution for $\e^{4A}$ and $\alpha$ at some fixed reference volume. Similarly, we define the fluctuating warped volume,
\begin{equation}
\mathcal{V}_\textrm{w}\left(\rho(x)\right) = \int \d^6 y \sqrt{\tilde g_6} \,\e^{-4A_0(y)} + \rho(x), \label{warped-modulus1}
\end{equation}
where we normalized the volume of the fiducial internal metric $\tilde g_{mn}$ such that $\int \d^6 y \sqrt{\tilde g_6}=1$. In addition to appearing in \eqref{warped-modulus2}--\eqref{warped-modulus1}, $x$-dependent fluctuations $\rho(x)$ also turn on off-diagonal metric components $g_{\mu n}$ \cite{Frey:2008xw}. However, since we are only interested in the scalar potential, we can restrict to constant $\rho(x)=\rho$ in the following and thus neglect this subtlety.

The no-scale property of \eqref{gkp} can also be seen directly from the effective scalar potential. Using \eqref{gkp} in \eqref{effaction2}, the potential simplifies to
\begin{align}
V &= \int \d^{6}y \sqrt{\tilde g_{6}}\, \frac{1}{\mathcal{V}_\textrm{w}^{2}} \bigg[ 16 (\tilde{\partial A})^2 + \e^{4A+\phi}|\tilde F_3|^2 \nl -\frac{1}{4}N_{\textrm{O}3}\, T_3 \e^{4A} \tilde \delta^{(6)} \bigg] + V_\textrm{np}.
\end{align}
Using \eqref{bianchi3} and integrating by parts, we then find that the terms in the brackets vanish,
\begin{align}
V = V_\textrm{np}. \label{potential-gkp}
\end{align}
Hence, as claimed above, the tree-level terms do not generate a potential for the volume modulus.

The large-volume limit of the compactification is obtained by taking $\rho \to \infty$, where $\rho$ is identified with the unwarped volume modulus \cite{Giddings:2005ff}. The localized sources are then far away from each other and the flux densities become small such that the part of the warp factor that varies over $y$ in \eqref{warped-modulus1} is negligible almost everywhere on the compact space. At large volumes, we thus find $\mathcal{V}_\textrm{w} \to \rho$. Note, however, that, although often claimed in the literature, the large-volume limit does in general not imply that \emph{derivatives} of the warp factor can be neglected in the equations of motion. Unless the D$3$ tadpole is cancelled pointwise on the compact space, they are of order of the fluxes at any volume, as can be read off from \eqref{bianchi3} (see \cite{Douglas:2010rt, Blaback:2010sj} for a discussion).

In order to determine the large-volume limit of \eqref{warped-modulus2} and \eqref{warped-modulus3}, we have to specify the region of the compact space we consider \cite{Giddings:2005ff}.
In weakly warped regions, $\rho$ dominates the right-hand sides of \eqref{warped-modulus2} and \eqref{warped-modulus3} such that their dependence on $\e^{-4A_0} = \alpha_0^{-1}$ can be neglected. In regions with very strong warping, on the other hand, $\rho$ can become negligible even at large volume, and \eqref{warped-modulus2} and \eqref{warped-modulus3} become volume-independent. This is true, e.g., sufficiently close to a brane, where the warp factor $\e^{-4A_0} = \alpha_0^{-1}$ diverges, or in warped throats like the KS solution, where the warp factor becomes exponentially large. Hence, we find the two limits
\begin{align}
& \textrm{\footnotesize weak:} && \mathcal{V}_\textrm{w} \to \rho,  && \e^{-4A} \to \rho,  && \alpha^{-1} \to \rho, \label{limit1} \\
& \textrm{\footnotesize strong:} && \mathcal{V}_\textrm{w} \to \rho,  && \e^{-4A} \to \e^{-4A_0},  && \alpha^{-1} \to \alpha^{-1}_0, \label{limit2}
\end{align}
where the first (second) line refers to the large-volume limit in a weakly (strongly) warped region of the compact space. Note that the large-volume limit of $\mathcal{V}_\textrm{w}$ is independent of the region since $\mathcal{V}_\textrm{w}$ is an integrated quantity.

\subsection{b. Adding D3-branes}

Let us now discuss what happens when we insert D$3$-branes into such a flux background \cite{Giddings:2001yu, Giddings:2005ff}. The right-hand sides of \eqref{bianchi} and \eqref{einstein} then receive additional contributions
\begin{align}
\tilde \nabla^m \e^{-8A} \partial_m \alpha &= \ldots + N_{\textrm{D}3} T_3 \tilde \delta^{(6)}, \label{bianchi1} \\
\tilde \nabla^m \e^{-8A} \partial_m \e^{4A} &= \ldots + N_{\textrm{D}3} T_3 \tilde \delta^{(6)}, \label{einstein1}
\end{align}
where the charge and energy densities of the branes source $\alpha$ and $\e^{4A}$, respectively. The presence of the D$3$-branes thus perturbs both fields, which we can write as
\begin{equation}
\e^{4A} = \e^{4A_\textrm{bg}} + \delta \e^{4A}, \quad \alpha =  \alpha_\textrm{bg} + \delta \alpha. \label{xyz}
\end{equation}
As explained above, by $\e^{4A_\textrm{bg}}$ and $\alpha_\textrm{bg}$ we denote the ``background'' fields, i.e., the solution for $\e^{4A}$ and $\alpha$ in the approximation where the backreaction of the $N_{\textrm{D}3}$ D$3$-branes is not taken into account but with flux numbers chosen such that the tadpole condition would be satisfied in their presence.

Since the charge of a D$3$-brane equals its tension, the source terms for $\e^{4A}$ and $\alpha$ enter with the same sign in \eqref{bianchi1} and \eqref{einstein1}. A simple guess for $\delta \alpha$ is therefore
\begin{equation}
\delta \alpha = \delta \e^{4A} = \e^{4A}-\e^{4A_\textrm{bg}}, \label{xyz2}
\end{equation}
where the last equality follows from our ansatz \eqref{xyz}. Quite remarkably, this simple ansatz already captures the full backreaction of the D$3$-branes on the flux background.

The reason for this nice behavior is that the perturbation \eqref{xyz2} preserves the property $\alpha=\e^{4A}$ of the background solution \eqref{gkp}. In other words, D$3$-branes are mutually BPS with the flux background \cite{Giddings:2001yu, Giddings:2005ff}. Physically, this can be understood from the fact that the fluxes carry D$3$-brane charge, which has the effect that, classically, gravitational and electromagnetic forces on a D$3$-brane exactly cancel out (see, e.g., \cite{Blaback:2010sj} for a discussion). The backreaction of a D$3$-brane on the bulk fields is therefore rather mild in the sense that it only affects the behavior of $\e^{4A}$ and $\alpha$ in the vicinity of the brane but preserves the general form of the solution \eqref{gkp}. Since the supergravity fields still satisfy \eqref{gkp} after adding a D$3$-brane, also the no-scale structure is preserved. The scalar potential for the volume modulus is therefore still given by \eqref{potential-gkp}.

\subsection{c. Adding anti-D3-branes}

Let us now insert $p$ anti-D$3$-branes into the BPS flux background. Since an anti-D$3$-brane carries minus the charge of a D$3$-brane, the BPS property of the background is then lost, and the mutual backreaction of the perturbed supergravity fields leads to a violation of the conditions \eqref{gkp}. It was argued in \cite{Gautason:2013} that the form fields in the throat region are then of the general form
\begin{align}
H &= \e^{\phi-4A} \star_6 \left({ \alpha F_3 + X_3 }\right), \label{formfields1} \\ F_5 &= - (1+\star_{10}) \e^{-4A} \star_6 \d \alpha, \label{formfields2}
\end{align}
where $X_3$ is a closed $3$-form. Furthermore, $\tilde R_6$ can in general be non-zero, and $\phi$ and $\alpha$ are functions of the internal coordinates that are a priori unknown. Sufficiently far away from the anti-branes, the fields are expected to approach the BPS solution \eqref{gkp}.

The anti-branes contribute to the right-hand sides of the Bianchi identity and the Einstein equation like
\begin{align}
\tilde \nabla^m \e^{-8A} \partial_m \alpha &= \ldots - p T_3 \tilde \delta^{(6)}, \label{bianchi2} \\
\tilde \nabla^m \e^{-8A} \partial_m \e^{4A} &= \ldots + p T_3 \tilde \delta^{(6)}. \label{einstein2}
\end{align}
Here, the dots indicate other sources for $\alpha$ and $\e^{4A}$ such as the $3$-form fluxes, O$3$-planes and D$3$-branes that are already present in the BPS background. Note, however, that the omitted terms now take a more complicated form than in \eqref{bianchi} and \eqref{einstein} since the supergravity fields do not satisfy \eqref{gkp} anymore.

It follows from \eqref{bianchi2} and \eqref{einstein2} that, just like a D$3$-brane, an anti-D$3$-brane perturbs $\e^{4A}$ and $\alpha$ in its vicinity, which we again write as
\begin{equation}
\e^{4A} = \e^{4A_\textrm{bg}} + \delta \e^{4A}, \quad \alpha =  \alpha_\textrm{bg} + \delta \alpha. \label{ansatz-alpha}
\end{equation}
Due to its negative charge, however, the anti-D$3$-brane sources $\e^{4A}$ and $\alpha$ with the opposite sign. A natural first guess for $\delta \alpha$ is therefore
\begin{equation}
\delta \alpha \overset{?}{=} - \delta \e^{4A} = -\e^{4A}+\e^{4A_\textrm{bg}}, \label{alpha-guess}
\end{equation}
where the second equality follows from \eqref{ansatz-alpha}. If we had inserted the anti-branes into a mutually BPS flux background (i.e., a background where the fluxes carry negative D$3$-brane charge), this ansatz would already capture the full backreaction on $\alpha$ (cf. \eqref{xyz2} and the subsequent discussion). However, the non-BPS configuration considered here is more involved. Let us therefore discuss in more detail the two terms on the right-hand side of \eqref{alpha-guess}. The appearance of the first term can be justified as follows. The warp factor $\e^{4A}$ is required to become singular near an anti-D$3$-brane in order that its Laplacian in \eqref{einstein2} yields the delta function term on the right-hand side. Since the Laplacian of $\alpha$ in \eqref{bianchi2} must yield the same delta function but with the opposite sign, we conclude that $\alpha$ contains a singular term that diverges like $-\e^{4A}$. The first term on the right-hand side of \eqref{alpha-guess} is therefore determined by the charge and tension of the anti-D$3$-branes. In addition, however, there can be non-singular perturbations of $\alpha$ that are smooth at the anti-brane position. These are not determined by the anti-branes themselves but depend on the background into which they are inserted. In a general non-BPS solution, there is no reason why they should amount to $\e^{4A_\textrm{bg}}$ as in \eqref{alpha-guess}. Instead, $\alpha$ may receive additional perturbations due to other bulk fields that are perturbed by the anti-branes and then backreact on $\alpha$.
Denoting these additional non-singular perturbations by $\delta \bar \alpha$, we write
\begin{equation}
\delta \alpha = -\e^{4A}+\e^{4A_\textrm{bg}} + \delta \bar \alpha.
\end{equation}
Substituting this into \eqref{ansatz-alpha}, we then find
\begin{equation}
\alpha = \alpha_\textrm{bg} + \e^{4A_\textrm{bg}} - \e^{4A} + \delta \bar \alpha. \label{alpha}
\end{equation}
This is our general ansatz for $\alpha$ after the insertion of the anti-D$3$-branes. Note that, for $\e^{4A} \to \e^{4A_\textrm{bg}}$ and $\delta \bar \alpha \to 0$ sufficiently far away from the anti-branes, $\alpha$ approaches the background solution, $\alpha \to \alpha_\textrm{bg}$.

Let us now discuss the dependence of the scalar potential \eqref{effaction2} on the volume modulus.
In the presence of the anti-branes, the equations of motion are perturbed in a (presumably) complicated way, and it is not obvious whether the warped volume modulus is still defined by \eqref{warped-modulus2}--\eqref{warped-modulus1} or also appears in other fields that are not sourced in the BPS solution. In order to obtain the exact definition, we would have to determine the consistent fluctuations of the 10d equations of motion as in \cite{Frey:2008xw}, which is difficult in the absence of the full solution.
For the purpose of this paper, however, it is sufficient to know the definitions in the large-volume limit, and we consider it reasonable to assume that \eqref{limit1} and \eqref{limit2} still hold if the BPS solution is perturbed by a small number of anti-branes.

Thus, we can infer the dependence of the scalar potential on the volume modulus from its dependence on $\mathcal{V}_\textrm{w}$, $\e^{4A}$ and $\alpha$.
Unlike in the probe approximation, the anti-brane term in \eqref{effaction2} does not directly contribute to the scalar potential since $\e^{4A} \to 0$ at the anti-brane position. However, it contributes through its backreaction on the bulk fields.
Using \eqref{formfields2} in \eqref{effaction2}, we first conclude that the terms involving $\tilde R_6$, $(\tilde{\partial A})^2$, $(\tilde{\partial \phi})^2$ and $|\tilde F^\textrm{int}_5|^2$ only depend on the volume through their common prefactor $\mathcal{V}_\textrm{w}^{-2}$.
Hence, in the large-volume limit \eqref{limit1}, they scale like $\rho^{-2}$.  Furthermore, the anti-branes source singular terms in the energy densities of the $3$-form fluxes.
In terms of powers of the warp factor, they diverge like
\begin{equation}
\e^{-\phi} |\tilde H|^2 \sim \e^{\phi} |\tilde F_3|^2 \sim \e^{-8A} \alpha(y_0)^2 \label{ed1}
\end{equation}
at the anti-brane position, as was shown in \cite{Gautason:2013} for fully localized anti-branes (for the simplified case of partially smeared anti-branes, this is consistent with the near-brane behavior found earlier in \cite{Bena:2012vz}). Hence, in the scalar potential \eqref{effaction2}, the flux terms scale like
\begin{equation}
\frac{1}{\mathcal{V}_\textrm{w}^{2}} \e^{-4A} \alpha^2. \label{fluxterms}
\end{equation}
Naively, one might now conclude that the large-volume limit of this scaling is given by \eqref{limit1} since, at large volume, warping effects are negligible almost everywhere on the compact space. If this were true, the $3$-form fluxes would scale like $\rho^{-3}$ in the scalar potential. However, recall that, in the BPS solution prior to inserting the anti-branes, the flux terms cancel with the O-plane and warp factor terms in \eqref{effaction2} such that the total scalar potential is zero. What contributes to the scalar potential after inserting the anti-branes is therefore the deviation of the flux terms from the BPS solution, which is significant only in the vicinity of the anti-branes.
Since the anti-branes are located in a strongly warped region of the compact space, the large-volume limit of \eqref{fluxterms} should then be given by \eqref{limit2}, which yields $ \mathcal{V}_\textrm{w}^{-2} \e^{-4A} \alpha^2 \sim \rho^{-2}$.
We should stress, however, that this limit only makes sense if the warping is strong enough that, even at large volume, $\rho$ is negligible in \eqref{warped-modulus2} and \eqref{warped-modulus3} over a sufficiently large region of the compact space (i.e., $\e^{-4A_0} \gg \rho \gg 1$, $\alpha_0^{-1} \gg \rho \gg 1$). In the KS throat, the warping can be made exponentially large such that $\rho$ can be much smaller than the warp factor while still being large enough to justify the large-volume limit.

Summarizing the above discussion, we find that, in the large-volume limit and assuming that the anti-branes are inserted into a strongly warped throat, the scalar potential \eqref{effaction2} takes the form
\begin{equation}
V(\rho) = \frac{D}{\rho^2} + V_\textrm{np}(\rho), \label{brpotential}
\end{equation}
where $D$ is a coefficient that will be determined in the next section.
Thus, the volume scaling of the uplift term reproduces the leading order behavior expected from the probe potential \eqref{potential-d3}. Let us again emphasize here that \eqref{brpotential} is corrected by a variety of sub-leading effects. Apart from corrections due to the backreaction of the non-perturbative effects, this includes warping and string corrections, whose systematics were discussed in \cite{DeWolfe:2002nn, Giddings:2005ff}.

\section{IV. Coefficient of the uplift term}

In order to fix the coefficient $D$ of the uplift term, we make use of the results presented in \cite{Gautason:2013}, where the classical scale invariance of type II supergravity was exploited to derive a relation between the cosmological constant and the boundary conditions at the anti-brane position, which holds on-shell. In 10d Einstein frame, the cosmological constant was found to take the form
\begin{equation}
\Lambda = \frac{1}{4\mathcal{V}_\textrm{w}} \left( \mathcal{L}_\textrm{DBI} + \mathcal{L}_\textrm{CS} \right) + \Lambda_\textrm{np}, \label{cc}
\end{equation}
where $\mathcal{L}_\textrm{DBI} = -pT_3\e^{4A(y_0)}$ and $\mathcal{L}_\textrm{CS} = - pT_3\alpha(y_0)$ are the on-shell integrands of the DBI and CS parts of the anti-D$3$-brane action (with $y_0$ the anti-brane position) and $\Lambda_\textrm{np}$ is a contribution due to the non-perturbative effects. Let us stress here that $\e^{4A(y_0)}$ and $\alpha(y_0)$ refer to the boundary conditions of the fields in the supergravity approximation. This may sound confusing since the supergravity solution is not expected to be trustworthy anymore near the anti-branes. However, as discussed above, the 4d scalar potential (and, accordingly, the cosmological constant) in the large-volume limit can still be reliably computed from the 10d supergravity solution even though string corrections may become relevant close to the localized sources (see Section 3.4 of \cite{Gautason:2013} for a more detailed discussion of this subtlety).

We should also note that a key assumption made in \cite{Gautason:2013} is that the anti-branes do not source non-normalizable modes violating the BPS conditions $\alpha=\e^{4A}$ and $X_3=0$. In other words, deviations from the BPS flux background were assumed to only be generated locally but vanish far away from the anti-branes. This assumption was also imposed in \cite{Bena:2009xk, Bena:2011hz, Bena:2011wh} to derive the linearized deformation of the KS background by partially smeared anti-D$3$-branes. Furthermore, corrections to the classical flux background due to the backreaction of the non-perturbative effects were assumed to be sub-leading in \cite{Gautason:2013}. Such corrections were studied in \cite{Koerber:2007xk, Baumann:2010sx, Heidenreich:2010ad, Dymarsky:2010mf}, and it would be interesting to systematically analyze their effect on the cosmological constant.

At large volume and going to 4d Einstein frame, \eqref{cc} implies that the on-shell scalar potential reads
\begin{equation}
V(\langle \rho\rangle) = \frac{1}{2\langle \rho\rangle^2} \left( \mathcal{L}_\textrm{DBI} + \mathcal{L}_\textrm{CS} \right) + \frac{2\Lambda_\textrm{np}}{\langle \rho\rangle}. \label{cc1}
\end{equation}
In order to derive \eqref{cc}, a certain combination of the 10d equations of motion with the external Einstein equation was used in \cite{Gautason:2013}. From the 4d point of view, this amounts to adding to the effective scalar potential a combination of moduli equations and, in particular, a multiple of the equation of motion for $\rho$. One can verify that the non-perturbative term $\Lambda_\textrm{np}$ in \eqref{cc}, which was left undetermined in \cite{Gautason:2013}, is therefore related to the on-shell value of the non-perturbative term in the scalar potential via $2\Lambda_\textrm{np}/\langle \rho\rangle = V_\textrm{np}(\langle \rho\rangle) + \frac{3}{4} V^\prime_\textrm{np}(\langle \rho\rangle)$, where $^\prime = \partial/\partial\ln \rho$. Furthermore, computing the $\rho$ equation from \eqref{brpotential} yields $V^\prime_\textrm{np}(\langle \rho\rangle) = 2 D/ \langle \rho\rangle^2 $. Using both equations in \eqref{cc1} and comparing with \eqref{brpotential}, we then find $D= - \left( \mathcal{L}_\textrm{DBI} + \mathcal{L}_\textrm{CS} \right) $, and, hence,
\begin{equation}
V(\rho) = - \frac{1}{\rho^2} \left( \mathcal{L}_\textrm{DBI} + \mathcal{L}_\textrm{CS} \right) + V_\textrm{np}(\rho). \label{cc2}
\end{equation}
We can now substitute $\mathcal{L}_\textrm{DBI} = -pT_3\e^{4A(y_0)}$ and $\mathcal{L}_\textrm{CS} = - pT_3\alpha(y_0)$ into \eqref{cc2} and use that $\e^{4A} \to 0$ at the anti-brane position to find
\begin{equation}
V(\rho) = \frac{pT_3\alpha(y_0)}{\rho^2} + V_\textrm{np}(\rho). \label{finalpotential}
\end{equation}
Hence, in order to allow for a dS extremum with $V_\textrm{np}<0$, $\alpha(y_0)$ must be positive \footnote{This corrects a small mistake in the argument of \cite{Gautason:2013}, where $\alpha(y_0)$ was found to be \emph{negative}. The reason for the disagreement is that, contrary to an assumption in \cite{Gautason:2013}, $V_\textrm{np}<0$ does not imply $\Lambda_\textrm{np} < 0$ but rather $\Lambda_\textrm{np} - \frac{3}{8}  \unexpanded{\langle\rho\rangle} V^\prime_\textrm{np}( \unexpanded{\langle\rho\rangle}) = \Lambda_\textrm{np} - \frac{3}{4} pT_3\alpha(y_0)/\unexpanded{\langle\rho\rangle} < 0$. The main conclusion $\alpha(y_0) \neq 0$, which was argued in \cite{Gautason:2013} to lead to a flux singularity, is unaffected by this subtlety.}. Note that a finite $\alpha$ at the anti-brane position is known to be directly related to the appearance of flux singularities \cite{Blaback:2011pn, Gautason:2013}.

In order to see whether the uplift term in \eqref{finalpotential} agrees with the probe potential \eqref{potential-d3}, we have to determine $\alpha(y_0)$. This can be done as follows. Using \eqref{alpha} together with $\e^{4A(y_0)} = 0$, we find
\begin{equation}
\alpha(y_0) = \alpha_\textrm{bg}(y_0) + \e^{4A_\textrm{bg}}(y_0) + \delta \bar \alpha(y_0). \label{fgdfsgt}
\end{equation}
As stated before, $\alpha_\textrm{bg}$ is defined as the background solution where the anti-brane backreaction is not taken into account but with flux numbers adjusted such that the tadpole condition is consistently satisfied. Before inserting the anti-branes, one finds $\alpha = \e^{4A} \propto \e^{-(8\pi K)/(3g_sM)}$ near the tip of the KS throat, where $K$ and $M$ are the quanta of the $3$-form fluxes \cite{Klebanov:2000hb}. For a large flux background that is perturbed by a small number of anti-branes, we can then set $\alpha_\textrm{bg}=\e^{4A_\textrm{bg}}$ in \eqref{fgdfsgt} up to corrections of order $p/M$.

The only unknown part of \eqref{fgdfsgt} is the term $\delta \bar \alpha(y_0)$. Since $\delta \bar \alpha$ measures the backreaction of the bulk fields on $\alpha$ due to their perturbation by the anti-branes, one may speculate that it is suppressed by a factor $p/M$ compared to $\alpha_\textrm{bg}$ and $\e^{4A_\textrm{bg}}$ (although this is not obvious given the non-linear dynamics involved) and thus becomes negligible in \eqref{fgdfsgt}. As we will see momentarily, we can reproduce the exact uplift term of the probe potential under this assumption. Let us therefore propose that, in the limit of large volume and a small number of anti-branes, the boundary condition for $\alpha$ at the anti-brane position is such that $\delta \bar \alpha(y_0)=0$, i.e.,
\begin{equation}
\alpha(y_0) \overset{!}{=} \alpha_\textrm{bg}(y_0) + \e^{4A_\textrm{bg}(y_0)}. \label{conj}
\end{equation}
This is the boundary condition advertized in the abstract of this note.

Using $\alpha_\textrm{bg}(y_0)=\e^{4A_\textrm{bg}(y_0)}$ in \eqref{conj} and substituting the result into \eqref{finalpotential}, we find
\begin{equation}
V(\rho) = \frac{2pT_3\e^{4A_\textrm{bg}(y_0)}}{\rho^2} + V_\textrm{np}(\rho).
\end{equation} 
Hence, as claimed above, the uplift term generated by the anti-brane backreaction at leading order exactly agrees with the probe potential \eqref{potential-d3}, including the famous factor of $2$ in the coefficient.
For smaller volumes and/or a larger number of anti-branes, sub-leading corrections become important, and the probe approximation is expected to get worse.

{Let us comment here on a related result in \cite{Dymarsky:2011pm}, where it was shown that the linearized solution proposed in \cite{Bena:2009xk} for (partially smeared) anti-D3-branes in the non-compact KS background has the expected ADM energy $2pT_3\e^{4A_\textrm{bg}(y_0)}$. This suggests that embedding this solution or a (yet to be found) non-linear and fully localized completion of it into a compact setting should, at large volumes, yield the correct uplift term in the scalar potential. The result of the present work can thus be viewed as complementary to the one in \cite{Dymarsky:2011pm}: while the latter shows that the full non-linear solution should approach the linearized one of \cite{Bena:2009xk} sufficiently far away from the anti-branes in order to have the expected amount of energy, here we use the same requirement to constrain the behavior of the solution in the region close to the anti-branes where their backreaction becomes highly non-linear.}

It would obviously be very interesting to test whether the boundary condition \eqref{conj} is satisfied in the appropriate limit of the backreacted 10d solution. Although the full solution for anti-D$3$-branes in the KS background is not known, it should be possible to perform such a test in the simplified setup with partially smeared anti-D$3$-branes, which was studied in \cite{McGuirk:2009xx, Bena:2009xk, Bena:2011hz, Bena:2011wh} and other works.
{Furthermore, analogous boundary conditions should appear} in toy models of anti-brane backreaction like the anti-D$6$-brane model discussed in \cite{Blaback:2011nz, Blaback:2011pn, Apruzzi:2013yva}.
As noted above, the appearance of flux singularities follows from universal scaling symmetries of the supergravity equations and is therefore largely insensitive to the details of the particular setups. We therefore expect that simplified models of anti-brane backreaction are a reliable testing ground for the full problem. A difficulty might be that the value of $\alpha(y_0)$ can in general not be fixed by a local analysis of the equations of motion near the anti-brane position but instead depends on global properties of the solutions. This was shown in \cite{Blaback:2011pn} for the anti-D$6$-brane model, and it is also true for partially smeared anti-D$3$-branes in the KS background, as can be verified using the results of \cite{Bena:2012vz}. 
Since analytic global solutions are hard to find in the absence of simplifying properties like the saturation of a BPS bound, it might be more promising to test \eqref{conj} using numerical simulations.

\section{V. Conclusion}

We argued in this note that the dynamics of the volume modulus in a warped flux compactification with backreacting anti-branes is in exact agreement with the physical intuition from the usual probe arguments, provided the near-brane boundary conditions of the bulk fluxes satisfy \eqref{conj}.
Naively, one could have expected that the singular backreaction of the anti-branes is too strong and thus modifies the coefficient and/or the volume scaling of the uplift term even at large volume. The scalar potential and, accordingly, the cosmological constant would then have differed from the probe estimate. If above boundary condition is satisfied, however, this is not the case, which suggests that the correct supergravity description of anti-branes in warped flux backgrounds is allowed to involve singular fluxes.
It would therefore be very interesting to explicitly test \eqref{conj} in the appropriate limit, at least for the simplified situation of partially smeared anti-D$3$-branes, {or to check whether an analogue of \eqref{conj} appears in} the anti-D$6$-brane toy model studied in \cite{Blaback:2011nz, Blaback:2011pn, Apruzzi:2013yva}.
An important extension of our work would also be to better understand the various corrections to the leading order potential, building on the results obtained here and in previous works on warped effective field theory.

It would finally be crucial to settle the question whether or not anti-branes in warped flux backgrounds lead to physically acceptable solutions, i.e., whether flux singularities are resolved in string theory. On the one hand, the existence of such solutions is motivated from the dual gauge theory point of view (see, e.g., \cite{Kachru:2002gs, Dymarsky:2013tna}). Since the flux singularities appear to be closely tied to a divergent warp factor \cite{Gautason:2013}, one possibility is that they are simply resolved along with the string theory resolution of the warp factor singularity. Furthermore, other possibilities such as NS$5$-brane polarization \cite{Kachru:2002gs} have not been excluded yet. It was also suggested that $\alpha^\prime$ corrections might play a role in the resolution of flux singularities \cite{Cottrell:2013asa}, but it is not obvious whether this can be shown in a trustworthy regime. On the other hand, it does not
appear to be possible to regulate flux singularities by means of physical cutoffs such as a horizon \cite{Bena:2012ek, Vanriet:2013}, a regularized charge profile \cite{Blaback:2011nz}, a D-brane polarization radius \cite{Bena:2012tx, Bena:2012vz}, or a curvature scale \cite{Gautason:2013, Buchel:2013dla}. These peculiar features make it less obvious \cite{Gubser:2000nd} that flux singularities are resolved by a UV completion of supergravity than for the standard warp factor singularity (see furthermore \cite{Blaback:2012nf, Bena:2014bxa} for stability arguments against singular fluxes). The answer how string theory resolves flux singularities is therefore likely to be interesting also from a conceptual point of view.
\\

\begin{acknowledgments}
The author thanks Alex Buchel, William Cottrell, Anatoly Dymarsky, Fridrik Freyr Gautason, Stefano Massai, Gary Shiu, Yoske Sumitomo, Henry Tye, Thomas Van Riet and Marco Zagermann for useful discussions. This work was supported in part by the HKRGC under Grant 604213.
\end{acknowledgments}

\bibliography{groups}

\end{document}